\documentclass[12pt]{article}

\usepackage[a4paper,text={17.5cm,24.0cm},centering]{geometry}
\usepackage{latexsym,graphicx,amsmath,amssymb,mathrsfs,bm}
\usepackage{fancybox,color,wrapfig,framed,mathtools}
\usepackage{tikz-cd}
\usepackage[breaklinks=true,colorlinks=true,linkcolor=blue,urlcolor=blue,citecolor=blue]{hyperref}

\def\ri{\texttt{i}}
\def\rd{{\rm d}}
\def\re{{\rm e}}
\def\fr{\mbox{$\frac{1}{2}$}}
\def\frr{\mbox{$\frac{1}{4}$}}

\def\qand{\quad\mbox{and}\quad}

\def\eps{{\varepsilon}}

\renewcommand{\theequation}{\arabic{section}.\arabic{equation}}

\begin{document}

\begin{center}
  \textbf{\Large Reappraisal of Whitham's 1967 theory for}\\[2mm]

\textbf{\Large  wave-meanflow interaction in shallow water}
\vspace{.75cm}

\textit{Thomas J. Bridges$\ ^1$ \& Daniel J. Ratliff$\ ^2$}
\vspace{.25cm}

\textit{$\ ^1$Department of Mathematics, University of Surrey, Guildford GU2 7XH, UK}
\vspace{0.15cm}

\textit{$\ ^2$Department of Mathematics, Physics and Electrical Engineering,\\
Northumbria University, Newcastle upon Tyne, NE1 8ST, UK}
\vspace{0.5cm}
\end{center}

\begin{framed}
\noindent{\bf Abstract.} The modulation equations for Stokes waves
in shallow water coupled to wave-generated meanflow, derived in
Whitham (1967), based on an averaged Lagrangian  are re-visited.
Firstly, it is shown that they can be recast into two coupled classical
shallow water equations, with modified gravity having the sign
of the Whitham index: ${\rm sign}(\omega_0''\omega_2)$.
Secondly, it is shown that the amplitude of the meanflow and amplitude
of the wave are, in general, independent.  Thirdly, the implications of the coalescing
characteristics, whose unfolding is associated with the Benjamin-Feir
instability, are studied.
\end{framed}

\section{Introduction}
\label{sec-intro-bousinesq-2d}

One of the first applications of Whitham modulation theory, based on an averaged Lagrangian, 
was to Stokes waves in shallow water coupled to meanflow \textsc{Whitham}~\cite{w67}.
Four coupled modulation equations were derived with two 
for the amplitude and mean velocity of the mean flow,
coupled to a pair of
equations for the amplitude and wavenumber of the wave component. 
The characteristics of the coupled problem and the unfolding of a double
characteristic, and change of characteristic type, 
 was shown to be associated with the onset of the Benjamin-Feir instability.
 Going forward, the approach outlined by Whitham has been widely used
 throughout the study of nonlinear waves to identify not only instabilities but
 to also understand the long-time evolution of such waves.
 The original paper, however, stops short of discussing this transition 
 as the amplitude of these Stokes waves increases which has been shown to move to lower
 depth for larger waves(c.f. \cite{lh78a,lh78b,do11,an14}). Moreover, there are a number of simplifying assumptions
 within the analysis that only hold in specific cases and may not hold in general.

We revisit this problem, starting with a brief review of the formulation
in \cite{w67}.  Our main new observation is that systems consisting of a mean flow
and a wave modulation can be recast into the form of coupled shallow water equations:
\begin{equation}\label{wmes-summary-introduction}
\begin{array}{rcl}
h_t + (hu)_x &=&  \mathcal{F}_1(H,V)_x \\[2mm]
u_t + uu_x + gh_x &=& \mathcal{F}_2(H,V)_x\\[2mm]
H_t + (VH)_x &=& \mathcal{F}_3(h,u)_x \\[2mm]
V_t + VV_x +g'H_x &=& \mathcal{F}_4(h,u)_x \,,
\end{array}
\end{equation}
where $h,u$ are the mean depth and mean horizontal velocity, and $H,V$ are the
amplitude and group velocity of the wave component. $g$ is the usual gravitational
constant and $g'$ is associated with the Whitham stability index
\begin{equation}\label{sign-g-prime-intro}
{\rm sign}(g') = {\rm sign}(\omega''\omega_2)\,,
\end{equation}
which governs the stability of the uncoupled wave component.  Explicit
expressions for the coupling terms, $\mathcal{F}_j$, $j=1,\ldots,4$, are given in \S\ref{sec-coupled-eqns} below.

While the 
appearance of a shallow water equation for the modulation of mean flow
is expected, it is surprising that the wave component modulation also has
the form of the classical shallow water equations.    With the coupling, the characteristic type is no longer determined by the
sign of $g'$. Our
second observation is to clarify the independent role of the amplitude of
the mean flow.  In many analyses the bulk variation of the flow 
and the wave amplitude are directly related, as in the original work of Whitham (e.g.\ \cite{w67} and Section 16.9 of \cite{whitham-book}), the Hasimoto-Ono
equation \cite{ho72}, and the Davey-Stewartson equation~\cite{ds74}, but other models, such as the 
Benney-Roskes system~\cite{br69}, retain this independence. Our analysis here justifies why they should be 
 treated separately, lending additional insight into why such models may 
 be more effective. Our third observation is to study further the implications of the coalescing characteristics within the water wave problem.

An outline of the paper is as follows.  First the derivation of the
averaged Lagrangian in \cite{w67} is reviewed in \S\ref{sec-avg-lagr}.
Then, in sequence, the SWEs for pure mean flow (\S\ref{sec-b-eqn}),
pure wave (\S\ref{sec-a-eqn}), and then coupled SWEs (\S\ref{sec-coupled-eqns})
are derived.  The coupled modulation equations are studied in
\S\ref{sec-amplitudes} and \S\ref{sec-characteristics}, with particular
attention to the emergence of coalescing characteristics. In
\S\ref{subsec-nonl-cont-char} and in the Concluding Remarks
(\S\ref{sec-cr}) some of the nonlinear implications associated
with coalescing characteristics are discussed.

\section{The averaged Lagrangian and reduction}
\setcounter{equation}{0}
\label{sec-avg-lagr}

Luke's variational principle \cite{luke}, in two space dimensions and
time, is
\[
\delta\int_{t_1}^{t_2}\int_{x_1}^{x_2} L(\eta,\phi)\,\rd x\rd t = 0 \,,
\]
for the unknown free surface $\eta(x,t)$ and velocity potential
$\phi(x,y,t)$, in an inviscid, irrotational, constant density fluid, with 
\begin{equation}\label{luke-density}
L = \int_{-h_0}^\eta \left(
\phi_t + \fr\Big(\phi_x^2+\phi_y^2\Big) + gy\right)\,\rd y \,,
\end{equation}
where $g$ is the gravitational constant and $h_0$ the still water depth.
That this variational principle delivers the governing equations
and boundary conditions for irrotational and inviscid water waves
is proved in \textsc{Luke}~\cite{luke}
and in \S13.2 of \textsc{Whitham}~\cite{whitham-book}.\footnote{Historical note: the paper of Luke immediately preceded W67
in the same issue of JFM, and both papers were submitted on the same date.}

The free surface is expressed in a Fourier series as
\begin{equation}\label{N-eta}
\eta(x,t) = N(\theta) = b + \sum_{m=1}^\infty a_m\cos(m\theta)\,,
\end{equation}
with $a_1:= a$, and
\begin{equation}\label{eta-def}
\theta = kx - \omega t\,.
\end{equation}
The parameters $a$ and $b$ will play important roles as the representative
amplitudes of the wave and mean flow respectively.
The Fourier expansion of $\phi(x,y,t)$ is
\begin{equation}\label{phi-exp}
\phi(x,y,t) = u x - \gamma  t + \widehat\phi(\theta,y)\,, \qquad \mbox{with} \qquad \widehat \phi(\theta,y) = \sum_{m=1}^\infty \frac{A_m}{m}{\rm cosh}\Big(mk(h_0+y)\Big)\sin(m\theta)\,.
\end{equation}
Substituting the form of $N$ and $\widehat\phi$ into $L$ gives
\begin{equation}\label{L-before-avg}
L = \left(\fr u^2-\gamma\right)(h_0+N) + \fr g N^2 -
(\omega-u  k) \int_{-h_0}^N \widehat\phi_\theta\,\rd y +
\int_{-h_0}^N \left(\fr k^2\widehat\phi_\theta^2 + \fr\widehat\phi_y^2\right)\,\rd y\,.
\end{equation}
In \cite{w67}, the Fourier expansion is carried to third order,
\[
N(\theta) = b + a \cos(\theta) 
+ a_2 \cos(2\theta) + a_3\cos(3\theta) + \cdots\,,
\]
and
\[
\begin{array}{rcl}
\widehat\phi(\theta,y) &=& A_1 {\rm cosh}\left(
k(h_0+y)\right)\sin(\theta)+\frac{1}{2} A_2 {\rm cosh}\left(
2k(h_0+y)\right)\sin(2\theta)\\[2mm]
&&\hspace{1.0cm} + \mbox{$\frac{1}{3}$}A_3 {\rm cosh}\left(
3k(h_0+y)\right)\sin(3\theta)+\cdots\,.
\end{array}
\]
They are substituted into (\ref{L-before-avg}) and averaged using
the standard averaging operator
\begin{equation}\label{f-avg}
\overline{f} := \frac{1}{2\pi}\int_0^{2\pi} f(\theta)\,\rd\theta\,,
\end{equation}
giving
\begin{equation}\label{Lbar}
\begin{array}{rcl}
\mathscr{L}:=\overline{L} &=& \left(\fr u ^2-\gamma\right)(h_0+b) +
\fr g b^2 + \frr g \big(
a^2+a_2^2+a_3^2\big) \\[2mm]
&&\displaystyle \hspace{.75cm} -
(\omega-u k) \overline{\int_{-h_0}^N \widehat\phi_\theta\,\rd y} 
+ \overline{\int_{-h_0}^N \left(\fr k^2\widehat\phi_\theta^2 + \fr\widehat\phi_y^2\right)\,\rd y}\,.
\end{array}
\end{equation}
\textsc{Whitham}~\cite{w67} gives a detailed account of solving for
 $A_1$, $A_2$, $A_3$, $a_2$ and
$a_3$, all in terms of $a$ and $b$.  Back substitution gives the
reduced Lagrangian
\begin{equation}\label{eqn-25-w67}
\mathscr{L}
= \big(\fr u^2-\gamma\big)h + \fr gb^2 + \fr E\left\{
1- \frac{(\omega-u k)^2}{gk{\rm tanh}(kh)}\right\} +
\fr E^2 \frac{k^2D_0}{g{\rm tanh}(kh_0)}+\cdots\,,
\end{equation}
where $h=h_0+b$ with $h_0$ the still water level, 
\begin{equation}\label{E-def}
E=\fr g a_1^2\equiv \fr g a^2\,,
\end{equation}
is the energy density and
\begin{equation}\label{D0-def}
D_0 = \frac{(9T_1^4-10T_1^2+9)}{9T_1^3}\,,\quad\mbox{with}\ T_1 = {\rm tanh}(kh_0)\,.
\end{equation}
The reduced Lagrangian (\ref{eqn-25-w67}) is equation (25) in \cite{w67}.
Expand $h=h_0+b$, under the assumption that $b \ll 1$ in the first term and the denominator in the third term, thereby generating the averaged Lagrangian
to leading order in $a$ and $b$
\begin{equation}\label{reduced-lagr}
\begin{array}{rcl}
\mathscr{L}(a,b,\omega,k,\gamma,u)
&=& \big(\fr u^2-\gamma\big)(h_0+b) + \fr gb^2 + 
+\fr k E \frac{(1-T_1^2)}{T_1}b \\[4mm]
&&\hspace{1.5cm} + \fr E\left\{
1- \frac{(\omega-u k)^2}{gkT_1}\right\} +
\fr E^2 \frac{k^2D_0}{g T_1}+\cdots\,,
\end{array}
\end{equation}
This is the averaged Lagrangian from which the modulation equations
are derived, based on 
$\mathscr{L}_\omega$, $\mathscr{L}_k$,
$\mathscr{L}_\gamma$, and $\mathscr{L}_u$, along with the constraints
$\mathscr{L}_b=0$ and $\mathscr{L}_a=0$.  This Lagrangian is valid up to
quadratic order in $E$ and $b$.  An interesting analysis of
the Lagrangian (\ref{Lbar}), valid for general finite-amplitude waves, is given in \textsc{Whitham}~\cite{w81}.

\section{Limiting cases of the averaged Lagrangian}
\setcounter{equation}{0}
In order to best demonstrate how the averaged Lagrangian results in a set of coupled set of shallow water equations, it is first illuminating to consider the limiting cases in which the mean flow and wave action are considered separately.  In particular, this section will demonstrate the process in which the classical modulation system of waves in the absence of mean flow can be recast into the desired shallow water with a suitable remapping of the modulation variables into one representing the group velocity and the other the energy density.

\subsection{Mean Flow without the wave}
\label{sec-b-eqn}

In this section we look at what the Whitham modulation equations generate
when the basic state is just pure mean flow; that is
\[
\eta(x,t) = b\qand \phi(x,y,t) = u x - \gamma t\,.
\]
In shallow water hydrodynamics this state is called a \emph{uniform flow}
with depth $h=h_0+b$ and mean horizontal velocity $u$.
Whitham called the parameters
in basic states of this form pseudo-frequencies ($\gamma$) and pseudo-wavenumbers ($u$).  However, they are mathematically of the same form as ordinary
frequencies, producing modulation equations of the same form (a discussion
is in \S 3.1 of \cite{br20}).
The only difference is that averaging is not required in this case.
The basic state can be substituted directly into (\ref{luke-density})
giving
\begin{equation}\label{L-avg-b-only}
\mathscr{L}(b,\gamma,u)
= \big(\fr u^2-\gamma\big)(h_0+b) + \fr g(h_0+b)^2 \,.
\end{equation}
The Whitham modulation equations are
\begin{equation}\label{wme-b-eqns}
\begin{array}{rcl}
&&\mathscr{L}_b  = 0 \\[2mm]
&&\displaystyle
\frac{\partial\ }{\partial x} \mathscr{L}_u-\frac{\partial\ }{\partial t} \mathscr{L}_\gamma=0\\[4mm]
&&\displaystyle\frac{\partial u}{\partial t}+\frac{\partial\gamma}{\partial x}=0\,,
\end{array}
\end{equation}
Substituting gives
\begin{equation}\label{mean-flow-eqns-1}
\begin{array}{rcl}
&& \fr u^2 + g(h_0+b) -\gamma= 0 \\[2mm]
&&\displaystyle\frac{\partial b}{\partial t}+
\frac{\partial\ }{\partial x} (u(h_0+b))=0\\[4mm]
&&\displaystyle\frac{\partial u}{\partial t}+\frac{\partial\gamma}{\partial x}=0\,.
\end{array}
\end{equation}
Substitute the first equation into the third, thereby eliminating
$\gamma$, and replace $h_0+b$ by
$h$ noting that $h_0$ is constant,
\begin{equation}\label{swes-b}
h_t + (u h)_x = 0 \qand u_t + uu_x + gh_x = 0\,.
\end{equation}
These equations are exactly
the classical shallow water equations for the mean depth $h$ and
mean velocity $u$, delivered by modulation of the uniform flow.
The first equation of (\ref{mean-flow-eqns-1}) shows that $\gamma$
can be interpreted as the total head, whereas $\mathscr{L}_u$ can
be interpreted as the mass flux.

\subsection{Wave without the mean flow}
\label{sec-a-eqn}

Now consider the averaged Lagrangian for the wave component only,
\begin{equation}\label{wnl-lagr}
\mathscr{L}(a,\omega,k) = G(\omega,k) a^2 + \fr \Gamma a^4 + \cdots\,.
\end{equation}
We will
develop the modulation theory for the abstract Lagrangian (\ref{wnl-lagr})
following \S15.1 in \cite{whitham-book} where (\ref{wnl-lagr}) is used
as a starting point,
and then transform to the shallow water equations.
For reference, $G$ and $\Gamma$ for (\ref{reduced-lagr}) are
\[
G(\omega,k) = \frr g\left(1-\frac{\omega^2}{gkT_1}\right)\qand
\Gamma =  \frac{gkD_0}{4T_1}\,,
\]
although we will not need these explicit expressions in the theory.

The Whitham modulation equations, analogous to (\ref{wme-b-eqns}) are
\begin{equation}\label{wme-a-eqns}
\begin{array}{rcl}
&&\mathscr{L}_a  = 0 \\[2mm]
&&\displaystyle
\frac{\partial\ }{\partial x} \mathscr{L}_k-\frac{\partial\ }{\partial t} \mathscr{L}_\omega =0\\[4mm]
&&\displaystyle\frac{\partial k}{\partial t}+\frac{\partial \omega}{\partial x}=0\,.
\end{array}
\end{equation}
Our main observation is that Whitham modulation equations (\ref{wme-a-eqns}),
for the weakly nonlinear Lagrangian (\ref{wnl-lagr}),
are exactly the shallow water equations with modified gravity,
\begin{equation}\label{swes-a}
H_t + (HV)_x = 0 \qand V_t + VV_x + g' H_x = 0 \,,
\end{equation}
with $V$ and $H$ defined by
\begin{equation}\label{V-H-def}
V = -\frac{G_k(\omega_0(k),k)}{G_\omega(\omega_0(k),k)}=  c_g\qand H = \big|G_\omega\big|a^2\,,
\end{equation}
where $c_g$ is the group velocity of the linear waves, and the
modified gravity
\begin{equation}\label{g-prime-wavecomponent}
g' = \frac{1}{\big|G_\omega\big|}\omega_0''\omega_2\quad
\Rightarrow\quad {\rm sign}(g') = {\rm sign}(\omega_0''\omega_2)\,,
\end{equation}
with $\omega_2$ the weakly nonlinear correction to the frequency (defined
below in terms of $\Gamma$).  

The frequency $\omega_0(k)$ is deduced from the dispersion relation of the
linear problem defined by $G(\omega_0(k),k)=0$, with the assumption
$G_\omega(\omega_0(k),k)\neq0$, required for the definition of $H$.

The familiar Whitham instability criterion
(e.g. \S 15.1 in \cite{whitham-book}) shows up in the shallow
water equations (\ref{swes-a}) in the modified gravity.  The shallow
water equations (\ref{swes-a}) are ill posed ($g'<0$) precisely when the Whitham
modulation equations are elliptic.

There are a number of subtleties in going from (\ref{wme-a-eqns})
to the shallow water equations (\ref{swes-a}), and the details are now
given.
The nonlinear dispersion relation is obtained from the first
equation, $\mathscr{L}_a=0$, in (\ref{wme-a-eqns}),
\begin{equation}\label{nonl-disp-rel}
G(\omega,k) + \Gamma a^2 = 0 \,.
\end{equation}
Writing the nonlinear dispersion relation in the conventional
form
\[
\omega = \omega_0(k) + \omega_2(k)a^2 + \cdots\,,
\]
and substituting into (\ref{nonl-disp-rel}) gives
\begin{equation}\label{Gamma-omega2}
G(\omega_0,k) =0\qand \omega_2 = -\frac{\Gamma}{G_\omega}\,.
\end{equation}
The second equation in (\ref{wme-a-eqns}) 
gives conservation of wave action
\begin{equation}\label{cwa-a}
0 =
\frac{\partial\ }{\partial x} \mathscr{L}_k-\frac{\partial\ }{\partial t} \mathscr{L}_\omega =
\frac{\partial\ }{\partial x} \big(G_k a^2\big)-\frac{\partial\ }{\partial t} \big(G_\omega a^2\big) \,,
\end{equation}
to leading order. Now substitute $a^2= H/\big|G_\omega\big|$ and multiply by -1
\begin{equation}\label{cwa-a-1}
\frac{\partial\ }{\partial t} \left(\frac{G_\omega}{\big|G_\omega\big|} H\right) +
\frac{\partial\ }{\partial x} \left(-\frac{G_k}{G_\omega} \frac{G_\omega}{\big|G_\omega\big|} H\right)=0\,.
\end{equation}
With the assumption $G_\omega(\omega_0(k),k)\neq0$ the ratio $G_\omega/\big|G_\omega\big|$
is constant and so with $V=-G_k/G_\omega$  this equation reduces to the first of (\ref{swes-a}).

Now multiply the third equation in (\ref{wme-a-eqns}) by $G_\omega$
\begin{equation}\label{cwaves-Gw}
G_\omega k_t +G_\omega \omega_x = 0\,.
\end{equation}
To transform the second term, differentiate the nonlinear
dispersion relation (\ref{nonl-disp-rel}) with respect to $x$
\[
G_\omega\omega_x + G_k k_x + (\Gamma a^2)_x = 0 \,.
\]
Substitute into (\ref{cwaves-Gw}), and using the relation between $\omega_2$ and $\Gamma$ in (\ref{Gamma-omega2}),
and so
\[
k_t + Vk_x + \frac{1}{G_\omega}\left( \frac{G_\omega}{\big|G_\omega\big|}\omega_2 H\right)_x = 0 \,.
\]
Now multiply by $dV/dk$
\begin{equation}\label{V-k-eqn}
\frac{dV}{dk}k_t + V\frac{dV}{dk}k_x + \frac{1}{G_\omega}\frac{dV}{dk}
\left( \frac{G_\omega}{\big|G_\omega\big|}\omega_2 H\right)_x = 0 \,.
\end{equation}
The first two terms simplify to $V_t+VV_x$.  To simplify the third
term requires some calculation. First, the expression for $dV/dk$ is
\begin{equation}\label{det-1}
\frac{dV}{dk} = \left(-\frac{G_k(\omega_0(k),k)}{G_\omega(\omega_0(k),k)}\right)_k = 
\frac{1}{G_\omega^3}{\rm det}\left[\begin{matrix}
G_{\omega\omega} & G_{\omega k} & G_\omega \\
G_{k\omega} & G_{kk} & G_k \\ G_\omega & G_k & 0 \end{matrix}
\right]\,.
\end{equation}
But we need to relate this to $\omega_0''(k)$ and that relationship
is obtained by differentiating the linear dispersion relation
$G(\omega_0(k),k)=0$ with respect to $k$ twice gives
\begin{equation}\label{det-2}
G_\omega^3\omega_0''(k) = {\rm det}\left[\begin{matrix}
G_{\omega\omega} & G_{\omega k} & G_\omega \\
G_{k\omega} & G_{kk} & G_k \\ G_\omega & G_k & 0 \end{matrix}
\right]\,.
\end{equation}
Comparing (\ref{det-1}) with (\ref{det-2}) gives
\[
\frac{dV}{dk} = \omega_0''(k)\,,
\]
and so
\begin{equation}\label{V-k-eqn-1}
V_t + VV_x + \frac{1}{G_\omega}\omega_0''(k)
\left( \frac{G_\omega}{\big|G_\omega\big|}\omega_2 H\right)_x = 0 \,.
\end{equation}
Noting that $G_\omega/\big|G_\omega\big|$ is constant and that
$\omega_2$ is treated as a constant at this order, gives the
second of the shallow water equations in (\ref{swes-a}).

The above result is general, showing that Whitham modulation
theory applied to any weakly nonlinear wave, with averaged Lagrangian
in the form (\ref{wnl-lagr}),
 leads to the shallow water equations.  
 Ironically, one can conclude from this result that the WMEs show that
the modulation of weakly nonlinear
deep water waves is governed by the shallow water equations albeit
with $g'<0$ and so ill-posed.  However, with the addition of surface
tension there are parameter values where $\omega_2\omega_0''(k)>0$ 
(see Figure 1 in \cite{dr77}).  In this case $g'>0$ and so the WMEs are
indeed a shallow water model for modulation of deep water capillary-gravity
Stokes waves.

Unfortunately, the shallow water form of the WMEs
does not appear to carry over to fully finite amplitude waves.
In \S7 of \textsc{Whitham}~\cite{w81} (see also \S15.2 in \cite{whitham-book}),
 it is shown, by taking
a Legendre transform of the averaged Lagrangian, that the Whitham modulation
equations can be cast in the form
\[
I_t + \frac{\partial\ }{\partial x} J(k,I) = 0 \qand
k_t + \frac{\partial\ }{\partial x} \omega (k,I) =0\,,
\]
where $J=H_k$ and $H(k,I)$ is the Hamiltonian function (Legendre
transform of $\mathscr{L}$).  These equations
are close but not exactly of the form of shallow water equations.

\section{Wave-meanflow coupling}
\setcounter{equation}{0}
\label{sec-coupled-eqns}

We have shown that modulation of mean flow on its own satisfies
 a shallow water equation
with conventional gravity (\ref{swes-b}) and modulation
of the wave component on its
own also satisfies a shallow water equation
 with a modified gravity (\ref{swes-a}).
  We now consider
the coupled problem treated in \cite{w67} and show that the modulation equations
there are coupled shallow water equations.

The coupled Whitham modulation equations are obtained from 
\begin{equation}\label{wme-plus-amp-eqns-2}
\begin{array}{rcl}
&&\mathscr{L}_b  = 0 \\[2mm]
&&\displaystyle
\frac{\partial\ }{\partial x} \mathscr{L}_u-\frac{\partial\ }{\partial t} \mathscr{L}_\gamma=0\\[4mm]
&&\displaystyle\frac{\partial u}{\partial t}+\frac{\partial\gamma}{\partial x}=0\,,
\end{array}
\end{equation}
and
\begin{equation}\label{wme-plus-amp-eqns-1}
\begin{array}{rcl}
&&\mathscr{L}_a  = 0 \\[2mm]
&&\displaystyle
\frac{\partial\ }{\partial x} \mathscr{L}_k -\frac{\partial\ }{\partial t} \mathscr{L}_\omega=0\\[4mm]
&&\displaystyle\frac{\partial k}{\partial t}+\frac{\partial\omega}{\partial x}=0\,,
\end{array}
\end{equation}
with the averaged Lagrangian given in (\ref{reduced-lagr}).  The derivatives
needed for the first three equations, to leading order, are
\begin{equation}\label{L-meanflow-derivs}
\begin{array}{rcl}
\mathscr{L}_b &=& \fr u^2- \gamma + gb + \fr kE\frac{(1-T_1^2)}{T_1} \\[2mm]
\mathscr{L}_\gamma &=& -(h_0+b) \\[2mm]
\mathscr{L}_u &=& u(h_0+b) -\frac{(\omega+u k)}{gT_1}E\,.
\end{array}
\end{equation}
Generalising the interpretation of the first equation in (\ref{mean-flow-eqns-1}),
setting $\mathscr{L}_b=0$ gives a generalisation of total head for the coupled
flow
\[
\gamma = gb + \fr u^2 +\fr kE\frac{(1-T_1^2)}{T_1}\,.
\]
Substitute this expression for $\gamma$
into the third equation in (\ref{wme-plus-amp-eqns-2}). 
The second and third equation in
(\ref{L-meanflow-derivs}) are inserted into
conservation of wave action.  Combining these two equations
gives the mean flow equations
forced by the wave, to leading order,
\begin{equation}\label{b-beta-swes}
\begin{array}{rcl}
b_t + (u( h_0+b))_x &=& - (E/c_0)_x \\[2mm]
u_t + uu_x + gb_x &=& - \big(B_0E/(h_0c_0)\big)_x\,,
\end{array}
\end{equation}
where
\begin{equation}\label{B0-def}
B_0 = c_g -\fr c_0 = \frac{\omega_0h_0}{2}\left( \frac{1-T_1^2}{T_1}\right)
\,,\quad c_0 = \frac{\omega_0}{k}\,.
\end{equation}

We now consider the second triad of equations in (\ref{wme-plus-amp-eqns-1}).
Firstly, solve $\mathscr{L}_a=0$ for $\omega$, taking the positive square root,
\begin{equation}\label{omega-nonl}
\omega = \omega_0 - u k + \frac{kB_0}{h_0} b + k^2\frac{D_0}{c_0}E + \cdots\,,\quad \omega_0(k)=\sqrt{gk{\rm tanh}(kh_0)}\,.
\end{equation}
This then allows one to write,
 to leading order,
\begin{equation}\label{L-waveflow-derivs}
\begin{array}{rcl}
\mathscr{L}_\omega &=& -\frac{(\omega-u k)}{gkT_1}E  = 
-\frac{E}{\omega_0}+\mathcal{O}(Eb,E^2)\\[2mm]
\mathscr{L}_k &=& \frac{(\omega-u k)}{gkT_1}\left\{u+
\frac{\omega_0'(\omega-uk)}{\omega_0}
\right\}E+\mathcal{O}(Eb,E^2) \approx c_g\frac{E}{\omega_0}  +\mathcal{O}(Eb,E^2)\,.
\end{array}
\end{equation}
and substitute into the third equation, which gives to leading order,
\begin{equation}\label{k-coupled-case}
k_t + c_g k_x + \left(k^2\frac{D_0}{c_0}  E\right)_x =
-\frac{\partial\ }{\partial x}\left(
\frac{kB_0}{h_0} b \right)-ku_x \,. 
\end{equation}
The second and third equations in (\ref{L-waveflow-derivs})
are inserted into conservation of wave action, 
\begin{equation}\label{cwa-coupled-case}
0 =
 \frac{\partial\ }{\partial x} \mathscr{L}_k -\frac{\partial\ }{\partial t}\mathscr{L}_\omega=
\frac{\partial }{\partial t}\left(\frac{E}{\omega_0}\right) + \frac{\partial\ }{\partial x}\left(
c_g \frac{E}{\omega_0}\right) =0 \,.
\end{equation}
The two equations (\ref{k-coupled-case}) and
(\ref{cwa-coupled-case}) can be put into the form of shallow water
equations by taking
\begin{equation}\label{V-H-def-coupled}
H = \frac{E}{\omega_0} \qand V = c_g:=u+\omega_0'(k)\,.
\end{equation}
Conservation of wave action (\ref{cwa-coupled-case}) then becomes
\begin{equation}\label{HV-coupled-eqn}
H_t + (VH)_x = 0\,.
\end{equation}
To transform the equation (\ref{k-coupled-case}) into
shallow water form,
multiply (\ref{k-coupled-case}) by $V_k=\omega_0''(k)$ and take $\omega_2$ to be the coefficient of
$a^2$ in (\ref{omega-nonl}). This sequence of computations transforms (\ref{k-coupled-case}) into
\begin{equation}\label{wmes-swes-Veqn}
V_t + VV_x +g'H_x = -kV_ku_x - V_k\frac{kB_0}{h_0} b_x -\bigg(\frac{kB_0}{h_0}\bigg)_k b V_x \,,
\end{equation}
to leading order.
Here terms of the form $(\cdot)V_x$ on the right hand side are neglected in
\cite{w67}
as it is argued that they are convective terms (this argument is discussed
in \cite{w67} as ``neglect of the $F\kappa_x$ term'' in equation (51)).
However, as we will show later in \S\ref{sec-characteristics},
this term contributes to the characteristics 
and influences the Benjamin-Feir stability boundary as the amplitude varies.
  The modified gravity is
\begin{equation}\label{g-prime-def}
g' = \frac{2\omega_0}{g}\,\omega_0''\omega_2 \quad
\Rightarrow\quad {\rm sign}(g') = {\rm sign}(\omega''_0\omega_2)\,.
\end{equation}
The sign of $g'$ no longer controls the characteristic type since
the equations are coupled.  Analysis of the characteristics of the
fully coupled problem are given in \S\ref{sec-characteristics} below.

To summarise, the Whitham modulation equations (\ref{b-beta-swes}),
(\ref{HV-coupled-eqn}) and
(\ref{wmes-swes-Veqn}) are coupled shallow water equations for
$H,V,b,u$,
\begin{equation}\label{wmes-summary}
\begin{array}{rcl}
b_t + (u( h_0+b))_x &=& - (kH)_x \\[2mm]
u_t + uu_x + gb_x &=& -\big(kB_0H/h_0\big)_x\\[2mm]
H_t + (VH)_x &=& 0 \\[2mm]
V_t + VV_x +g'H_x &=& - kV_ku_x - V_k\frac{k B_0}{h_0}b_x -\bigg(\frac{kB_0}{h_0}\bigg)_k b V_x\,.
\end{array}
\end{equation}
The first pair are mean-flow equations forced by the wave component,
and the second pair are wave component equations forced by the mean-flow.

Whitham simplifies these equations further by
assuming that all the modulation parameters can be expressed in terms
of $a^2$ as
\begin{equation}\label{approximations}
b = \mathcal{O}(a^2)\,,\quad \gamma =\mathcal{O}(a^2)\,,\qand
 u = \mathcal{O}(a^2)\,.
\end{equation}
The argument for the approximation on $b$ is discussed in \S\ref{sec-amplitudes} below.
With the approximations (\ref{approximations})
the modulation equations (\ref{wmes-summary})
simplify to
\begin{equation}\label{wmes-summary-1}
\begin{array}{rcl}
b_t + h_0u_x &=& - k H_x  \\[2mm]
u_t + gb_x &=& -\frac{kB_0}{h_0}H_x - \frac{(kB_0)_k}{h_0V_k}HV_x\\[2mm]
H_t +c_gH_x + HV_x &=& 0 \\[2mm]
V_t +c_gV_x + g'H_x &=& - k V_k u_x - V_k\frac{kB_0}{h_0} b_x  \,.
\end{array}
\end{equation}
The last term in the second equation can be neglected at
this order as it does not affect the characteristic calculations in
\S\ref{sec-characteristics}
 but it is retained for comparison as Whitham includes it. The $c_g$ on the
left-hand side of the second pair is the constant $c_g$ of the basic
state.  The equations (\ref{wmes-summary-1})
are equivalent, to equations (51)-(54) in
\cite{w67}.

\subsection{Independence of the amplitude parameters $b$ and $a$}
\label{sec-amplitudes}

The amplitude parameters $a$ and $b$ are independent small parameters.
This can be seen by re-arranging the equations $\mathscr{L}_a=0$ and
$\mathscr{L}_b=0$.  Re-arranging $\mathscr{L}_b=0$
in (\ref{L-meanflow-derivs}) gives
\[
 gb + \frac{gk}{4}\left(\frac{1}{T_1}-T_1\right)a^2 = \gamma - \fr u^2 \,.
\]
Similarly, re-arranging $\mathscr{L}_a=0$ and taking the square root
gives, to leading order
\[
\frac{\omega}{\omega_0} = 1 + \frac{D_0}{2T_1}(ka)^2 + \frac{1}{2}
\left(\frac{1}{T_1}-T_1\right) kb + \frac{ku}{\omega_0}\,.
\]
Combining into one equation
\begin{equation}\label{a-b-eqn}
\left[ \begin{matrix} \frac{D_0}{2T_1} & \frac{1}{2}\left( \frac{1}{T_1}-T_1\right) \\[2mm] \frac{1}{2}\left( \frac{1}{T_1}-T_1\right) & 2 \end{matrix}\right]
\begin{pmatrix} (ka)^2 \\[2mm] kb \end{pmatrix} =
\begin{pmatrix} \frac{\omega}{\omega_0}-1-\frac{u}{c_0}\\[2mm]
\frac{2\gamma k}{g}-\frac{ku^2}{g}\end{pmatrix}\,.
\end{equation}
$(kb)$ and $(ka)^2$ are independent since the determinant is
\[
{\rm det}
\left[ \begin{matrix} \frac{D_0}{2T_1} & \frac{1}{2}\left( \frac{1}{T_1}-T_1\right) \\ \frac{1}{2}\left( \frac{1}{T_1}-T_1\right) & 2 \end{matrix}\right] =
\frac{1}{8T_1^4}\left( 9 - 12 T_1^2+13T_1^4-2T_1^6\right)\,.
\]
and it is nonzero for $0\leq T_1^2\leq 1$.
The equation (\ref{a-b-eqn}) can be solved uniquely for $a$ and $b$ as functions of
the modulation parameters
\begin{equation}\label{ba-parameters}
a := a(\omega,k,\gamma,u) \qand b := b(\omega,k,\gamma,u)\,.
\end{equation}
These expressions for $a$ and $b$ can then be substituted into the reduced
Lagrangian (\ref{wnl-lagr}), giving a Lagrangian dependent only on the
modulation parameters
\[
\mathcal{L}(\omega,k,\gamma,u) = \mathscr{L}(a(\omega,k,\gamma,u),b(\omega,k,\gamma,u),\omega,k,\gamma,u)\,.
\]
The Whitham modulation equations then reduce to the closed system
\[
k_t+\omega_x = 0\,,\quad (\mathcal{L}_k)_x-(\mathcal{L}_\omega)=0\,,\quad
u_t+\gamma_x= 0\,,\quad(\mathcal{L}_u)_x- (\mathcal{L}_\gamma)=0\,.
\]
These equations can be much more complicated
and more difficult to
work with, than the equivalent equations that include
the amplitudes.  However, this approach -- elimination of the
amplitudes -- is the more useful choice in some examples
(e.g.\ \cite{r17,br19,br20}).

The above result highlights definitively that the amplitude parameters $a$ and $b$ are in general independent.
However, implicitly in \cite{w67} and explicitly in \S16.9 of \cite{whitham-book} the two parameters are taken to be related. The relationship between
$b$ and $a^2$ is
not true in general as shown above, but in equation (16.99) of
\cite{whitham-book}, with additional assumptions, a relationship 
between $b$ and $a$ of the following form is derived
\begin{equation}\label{ba-relationship}
kb = -\frac{1}{2kh_0(1-c_g^2/gh_0)}\left(\frac{2c_g}{c_0}-\frac{1}{2}\right)
k^2a^2\,.
\end{equation}
This relationship induces a constraint on modulation space.  To see
this, substitute (\ref{ba-relationship}) into (\ref{ba-parameters})
and eliminate $b$ and $a$.  The result is
a constraint between the parameters,
\[
F(\omega,k,\gamma,u) = 0\,.
\]
This constraint defines a hypersurface in parameter space, and restricts the
equations (\ref{wmes-summary}) to be valid only on the hypersurface. 
This restriction may be valid in the weakly nonlinear limit,
but it does not carry over to higher order. More importantly it is not
necessary.

\section{Comparison with coupled NLS equations}
\setcounter{equation}{0}
\label{sec-cnls}

We digress briefly to give another point of view on the
 ``independence of amplitudes'' question, using
the modulation of the two-phase wavetrains of coupled
nonlinear Schr\"odinger (CNLS) equations. 
Take the CNLS equations in the following form     
\begin{equation}\label{cnls}
  \begin{array}{rcl}
\displaystyle    2\ri\frac{\partial \Psi_1}{\partial t} +
    \alpha_1 \frac{\partial^2 \Psi_1}{\partial x^2} +
    (\beta_{11}|\Psi_1|^2 + \beta_{12}|\Psi_2|^2)\Psi_1 &=& 0 \\[4mm]
    \displaystyle    2\ri\frac{\partial \Psi_2}{\partial t} +
    \alpha_2 \frac{\partial^2 \Psi_2}{\partial x^2} +
    (\beta_{12}|\Psi_1|^2 + \beta_{22}|\Psi_2|^2)\Psi_2 &=& 0\,,
  \end{array}
  \end{equation}
where $\Psi_1(x,t)$ and $\Psi_2(x,t)$ are complex-valued, and
$\alpha_1,\alpha_2$ and $\beta_{11},\beta_{12},\beta_{22}$ are 
nonzero real parameters with $\beta_{11}\beta_{22}-\beta_{12}^2\neq0$.
More general versions of CNLS exist, but this model will be sufficient
to clarify our argument.

Consider the basic two-phase wavetrain solution,
\begin{equation}\label{2-phase}
\Psi_1(x,t) = A_1 \re^{\ri(k_1x-\omega_1t)} \qand \Psi_2(x,t) = A_2
\re^{\ri(k_2x-\omega_2t)}\,.
\end{equation}
Substitution into (\ref{cnls}) gives
the nonlinear dispersion relations
\begin{equation}\label{nonl-disp-cnls}
2\omega_1 =\alpha_1 k_1^2 - \beta_{11}|A_1|^2 - \beta_{12}|A_2|^2\qand
2\omega_2 =\alpha_2 k_2^2 - \beta_{12}|A_1|^2 - \beta_{22}|A_2|^2\,.
\end{equation}
Rewrite as a matrix-vector equation to highlight the independence
of the amplitudes
\[
\left[\begin{matrix} \beta_{11} & \beta_{12} \\ \beta_{12} & 
\beta_{22} \end{matrix}\right] \begin{pmatrix} |A_1|^2 \\
|A_2|^2 \end{pmatrix} = \begin{pmatrix} \alpha_1k_1^2-\omega_1 \\ \alpha_2k_2^2-\omega_2 \end{pmatrix}\,.
\]
This equation is the analogue of (\ref{a-b-eqn}).
With $\beta_{11}\beta_{22}-\beta_{12}^2\neq0$, it is clear that the
amplitudes $|A_1|^2$ and $|A_2|^2$ are independent,
 and if a relationship between the amplitudes,
$|A_2|=\sigma |A_1|$, exists 
then a constraint on wavenumber-frequency space
would be introduced,
\begin{equation}\label{cnls-constraint}
(\beta_{12}+\beta_{22}\sigma)(\alpha_1k_1^2-\omega_1) =
(\beta_{11}+\beta_{12}\sigma)(\alpha_2k_2^2-\omega_2)\,.
\end{equation}
This constraint is a hypersurface in $(\omega_1,\omega_2,k_1,k_2)$ space,
which will not be satisfied in general. Moreover, this constraint would
restrict the admissable parameter values in the modulation of the
two-phase wavetrain (see below for WMEs).

Another reason this CNLS example is interesting is that
the WMEs, modulating the two-phase wavetrain (\ref{2-phase}),
 are in fact equivalent to the coupled SWEs,
\begin{equation}\label{wmes-summary-cnls}
\begin{array}{rcl}
\displaystyle
\frac{\partial h_1}{\partial t} + \frac{\partial (h_1u_1)}{\partial x} &=& 0 \\[4mm]
\displaystyle\frac{\partial u_1}{\partial t} + u_1\frac{\partial u_1}{\partial x} +
g_1\frac{\partial h_1}{\partial x} &=& \displaystyle\alpha_1\beta_{12}\frac{\partial h_2}{\partial x}\\[4mm]
\displaystyle\frac{\partial h_2}{\partial t} + \frac{\partial (h_2u_2)}{\partial x} &=& 0 \\[4mm]
\displaystyle\frac{\partial u_2}{\partial t} + u_2\frac{\partial u_2}{\partial x} +
g_2\frac{\partial h_2}{\partial x} &=& \displaystyle\alpha_2\beta_{12}
\frac{\partial h_1}{\partial x} \,.
\end{array}
\end{equation}
with $g_1=-\alpha_1\beta_{11}$ and $g_2=-\alpha_2\beta_{22}$.
These equations have almost the same form as the modulation equations in
W67, with the main difference being the form of the
coupling terms.  
A derivation of the modulation equations (\ref{wmes-summary-cnls})
is given in Appendix \ref{app-a}.

\section{Coalescing characteristics}
\setcounter{equation}{0}
\label{sec-characteristics}

The main application in W67 is to show that the system (\ref{wmes-summary-1})
has coalescing characteristics whose unfolding captures the Benjamin-Feir instability and its transition. Here
we review that theory and discuss the nonlinear implications.

The system (\ref{wmes-summary-1}) is a system of first-order PDEs
\begin{equation}\label{U-vector-eqn}
{\bf U}_t + {\bf M}({\bf U}) {\bf U}_x = {\bf 0}\,,
\end{equation}
with
\begin{equation}\label{M-def}
{\bf U} = \begin{pmatrix} b\\ u\\ H\\ V\end{pmatrix}
\qand
{\bf M} = \left[ \begin{matrix}
u & h_0 & k & 0 \\
g & u & \frac{kB_0}{h_0} & \frac{(kB_0)_k}{h_0V_k}H  \\
0 & 0 & c_g & H \\ 
\frac{kB_0}{h_0}V_k & kV_k & g' & c_g+\frac{(kB_0)_k}{h_0} b \end{matrix}\right]\,.
\end{equation}
The characteristics, denoted by $C$, are the eigenvalues of ${\bf M}({\bf U})$ evaluated at a given state, which we can choose as ${\bf U}_0 = (b,0,H,c_g)$ without loss of generality since the fluid velocity can be removed via a suitable Gallilean shift. These characteristics satisfy the quartic equation
\begin{equation}\label{C-quartic}
\begin{array}{rcl}
\Delta(C,k,E,b) &:=&\Delta(C;a,b):=-\big( gh_0 + 2B\big)\\[2mm]
&=&(C^2-gh_0)\big[ (C-c_g)^2 -g'H\big]\\[2mm]
&&
-\left( gh_0 + 2B_0C + B_0^2\right) \frac{k^2 V_k}{h_0}H- (C+B_0)(kB_0)_k \frac{k}{h_0}(C-c_g)H\\[2mm]
&&
- (C-c_g)\bigg[(c^2-gh)\frac{(kB_0)_k}{h_0}+g(C-c_g)\bigg]b+\mathcal{O}(H^2,b^2,bH)= 0\,.
\end{array}
\end{equation}
This equation is exact to the order of approximation of the averaged Lagrangian
(\ref{wnl-lagr}).  
We are interested in the type (elliptic or hyperbolic) of the characteristics.
The amplitude parameter $b$ appears to the same order as the energy density $H$,
as expected. Hence, we expect the mean level plays a role in the stability and
 characteristic speeds of the wave to some order in the analysis.

A complete analysis of the roots of this quartic requires computation.
However, Whitham notes that in the limit as the amplitude of the wave goes to
zero (limit $H\to0$ here) two of the characteristics coalesce. When $H=0$,
the characteristics are
\[
C = \pm \sqrt{gh_0} \qand C = c_g\ \mbox{(multiplicity 2)}\,.
\]
The first two characteristics are the shallow water modes, and will remain
real for $H$ small.  The second two are coalesced characteristics.  
These latter characteristics can become
complex, thereby making (\ref{U-vector-eqn}) elliptic,
 when $H$ is perturbed away from zero.

To unfold the double characteristic let
\begin{equation}\label{C-H-def}
C = c_g + \Upsilon \sqrt{H} + \mathcal{O}(H)\,,
\end{equation}
and substitute into (\ref{C-quartic})
\begin{equation}\label{C-quartic-Lambda}
(c_g^2-gh_0)\big[ \Upsilon^2H -g'H\big]
-\left( gh_0 + 2B_0c_g + B_0^2\right) \frac{k^2 V_k}{h_0}H=\mathcal{O}(H^{3/2})\,.
\end{equation}
Equating the $\mathcal{O}(H)$ terms to zero gives
\begin{equation}\label{Upsilon-def}
\Upsilon^2 = g' - \frac{(gh_0 + 2B_0c_g + B_0^2)}{gh_0-c_g^2} \frac{k^2 V_k}{h_0} =  \frac{k^2\omega_0''(k)}{h_0}\bigg(kh_0 D_0- \frac{(gh_0 + 2B_0c_g + B_0^2)}{gh_0-c_g^2}\bigg)
\end{equation}
which agrees with equation (57) in \cite{w67} noting
that $g'=k^3D_0V_k$ from (\ref{g-prime-def}).
Whitham computes the sign of the right-hand side of (\ref{Upsilon-def})
and shows that
it is negative when $kh_0$ is greater than $\approx 1.363$, which is
the familiar Benjamin-Feir instability threshold \cite{benjamin67,bm95}.
Hence for $H\ll1$ the double characteristic at $C=c_g$ splits into two
\begin{equation}\label{char-splitting}
C = c_g \pm \sqrt{\omega_0''(k)\omega_2^{eff}(k) H} + \mathcal{O}(H)\,, \quad \mbox{where} \quad \omega_2^{eff} = \frac{k^2}{h_0}\bigg(kh_0 D_0- \frac{(gh_0 + 2B_0c_g + B_0^2)}{gh_0-c_g^2}\bigg)
\end{equation}

The link of the elliptic characteristic type with instability can be seen
 by looking at the
eigenvalues of the linearised time dependent problem associated with
(\ref{U-vector-eqn}),
\[
{\bf U}_t + {\bf M}({\bf U}_0){\bf U}_x = 0\,,
\]
with a spectral ansatz
\[
{\bf U}(x,t) = \widehat{\bf U} \re^{\lambda t + \ri \alpha x}\,,
\]
where $\alpha$ the wavenumber of the perturbation, the eigenvalues are
\begin{equation}\label{lambda-def}
\lambda_j = -\ri C_j \alpha\,,\quad j=1,2,3,4\,,
\end{equation}
where $C_j$, $j=1,2,3,4$, are the four characteristics.  Let
\[
C_1 = -\sqrt{gh_0} + \mathcal{O}(H)\,,\quad 
C_2 = +\sqrt{gh_0} + \mathcal{O}(H)\,,
\]
and
\[
C_3 = c_g + \Upsilon\sqrt{H} + \mathcal{O}(H)\,,\quad
C_4 = c_g - \Upsilon\sqrt{H} + \mathcal{O}(H)\,,
\]
with $\Upsilon$ the positive square root of (\ref{Upsilon-def}).

The change of $\Upsilon$ from real to complex is reflected
in the eigenvalues in (\ref{lambda-def}) becoming a complex quartet.
The position of the four eigenvalues is shown in Figure
 \ref{fig_eigenvalues} below.  The outer eigenvalues are the simple
eigenvalues $\pm\ri\sqrt{gh_0}\alpha$.  The inner eigenvalues are
the complex quartet.  They are two double eigenvalues at $H=0$ and
then in the unfolding ($kh_0$ moved away from $1.363$) they become
complex for $kh_0>1.363$. This figure
is consistent with Figure 2 in \cite{bm95}.
\begin{figure}[ht]
\begin{center}
\includegraphics[width=6.5cm]{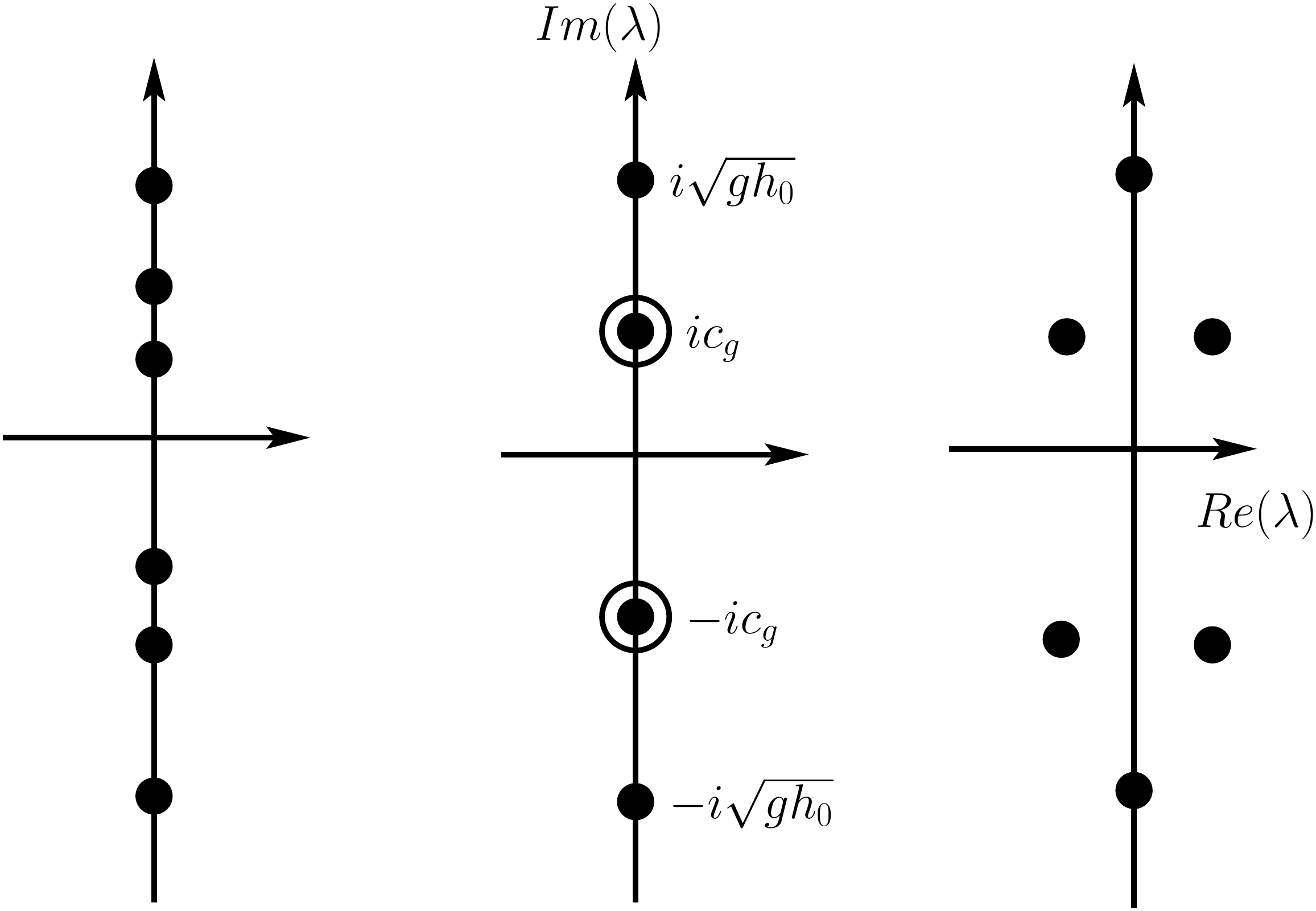}
\end{center}
\caption{Collision of purely imaginary eigenvalues in the Whitham equations,
with $\alpha=1$. The left picture corresponds to $kh_0<1.363$ and the
right picture corresponds to $kh_0>1.363$.  The middle picture shows
the case $H=0$.}
\label{fig_eigenvalues}
\end{figure}

\section{Nonlinear continuation of coalescing characteristics}
\label{subsec-nonl-cont-char}

\begin{figure}[ht]
\begin{center}
\includegraphics[width=6.5cm]{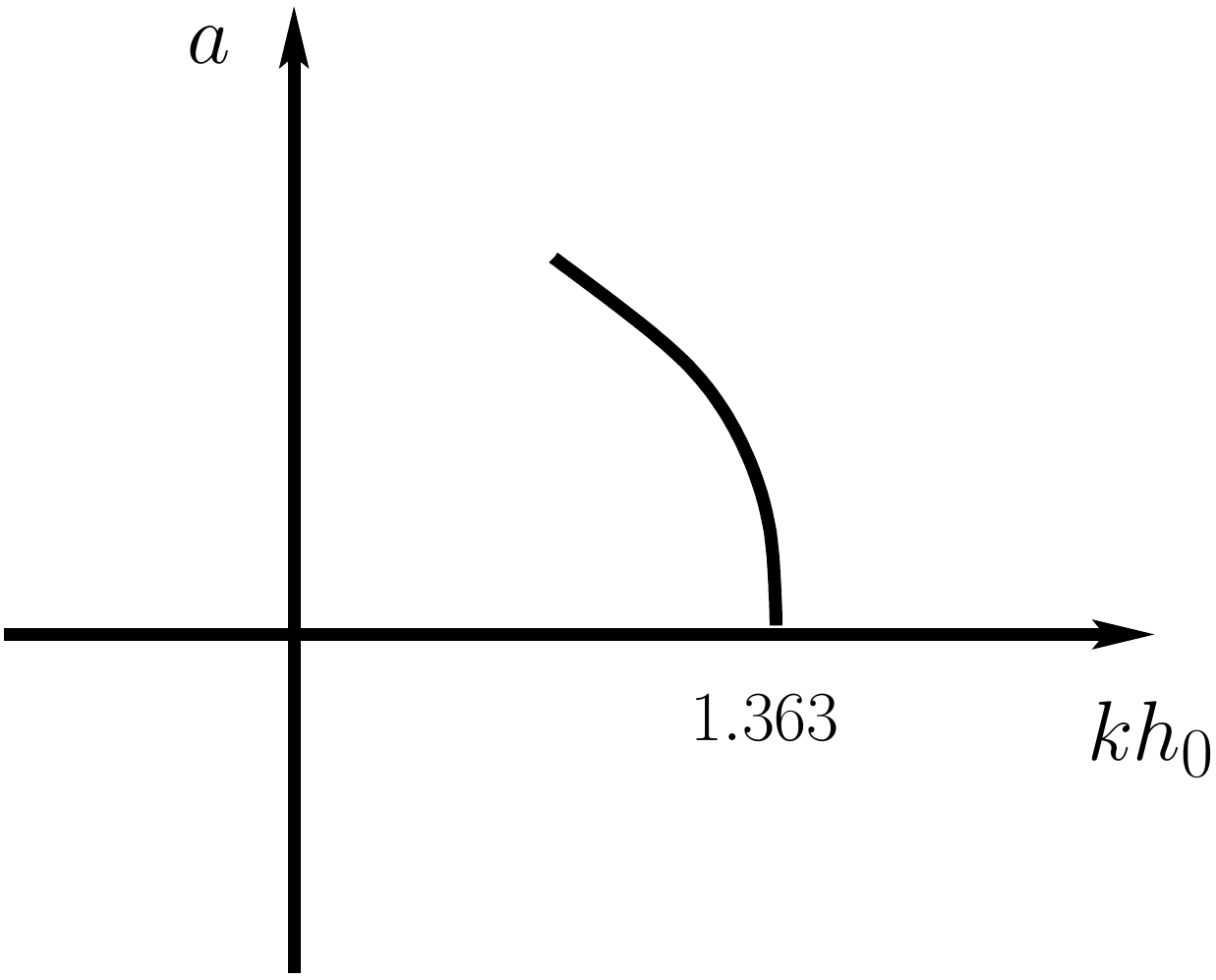}
\end{center}
\caption{Curve along which ${\rm det}[{\bf M}(U_0)-C{\bf I}]$ has a double
characteristic.}
\label{fig_kh-a-diagram}
\end{figure}
In the previous section we looked at the effect of amplitude on the unfolding
of the coalesced characteristics with $kh_0$ fixed.  In this section we enforce
 the coalescence and
look for a curve in $(kh_0,a)$ space where the coalesence is continued to
finite amplitude.  In this case it is
necessary to solve the ``coalescing characteristic equations''
\[
\Delta(C;a,k) = 0 \qand \frac{\partial\ }{\partial C}\Delta(C;a,k) = 0 \,.
\]
These two equations define a curve in the $(kh_0,a)$ plane.  At leading
order, this curve is a parabola emerging from the point
$(kh_0,a)= ((kh_0)^{\tiny crit},0)$ with $(kh_0)^{\tiny crit} =1.363$.
The leading order form of this curve can be obtained by expanding
$C$, $k$, and $E$ in terms of the amplitude $a$, resulting in the coalescing characteristic branch
\begin{equation}\label{C-kh-curve}
C = c_g + \frac{1}{2} \chi ga^2  + \frac{1}{2}\frac{(kB_0)_k}{h_0} b +\mathscr{O}(a^3,b,ab)\,\,,
\end{equation}
and 
\[
kh_0 = (kh_0)^{\tiny crit} -\frac{\chi^2}{2\omega_0'' (\omega_2^{eff})_k} +\xi b+ \mathscr{O}(a^3)\,,
\]
where
\[
\chi = -\frac{k(B_0+c_g)}{2h_0(gh_0-c_g^2)}\bigg((kB_0)_k+\frac{k \omega_0''(B_0c_g+gh_0)}{gh_0-c_g^2}\bigg)\,,
\]
and
\[
\xi = \frac{k}{h_0^2(gh_0-c_g^2)(\omega_2^{eff})_{h_0}}\left[\frac{(c_g + B_0)(kB_0)_k^2}{2h_0\omega_0''} + \frac{k(B_0c_g + gh_0)}{gh_0-c_g^2}\big((B_0c_g+gh_0)-(B_0 + c_g )(kB_0)_k \big))\right]\,.
\]
It becomes apparent that, to the order of the Lagrangian presented here, that the  Benjamin-Feir 
stability boundary is (incorrectly) increasing with amplitude. This is rectified by
including the higher order Stokes frequency correction of order $a^4$ given 
by $\omega_4^{eff}>0$,  which includes the effects of mean flow:
\[
kh_0 = (kh_0)^{\tiny crit} -\frac{1}{2}\bigg(\omega_4^{eff}+\frac{\chi^2}{\omega_0''(k)}\bigg) \frac{gh_0a^2}{(\omega_2^{eff})_k} +\xi b+ \mathscr{O}(a^3)
\]
which then gives the stability threshold as a decreasing function of wave amplitude, in 
agreement with the results of other works~\cite{km83,s05}, and is shown
schematically in Figure \ref{fig_kh-a-diagram}. The role of the bulk variations
are encoded into this higher order correction using (\ref{ba-relationship}) and an open question remains as to
how the mean variation $b$ contributes to this term, which could be discerned via a 
higher order analysis of the Lagrangian using the notions of this paper but is outside the scope of the work presented here.
At each point on the curve there is a double characteristic. This is just 
the starting point for the curve.  At finite amplitude there is the
possibility of bifurcations in the $(kh_0,a)$ plane, producing
multiple coalescing characteristics.  Even simple problems like the coupled nonlinear Schr\"odinger
equations have more than one set of coalescing characteristics at some
parameter values
(e.g.\ Figure 5 in \S5 of \cite{br19}).

\section{Concluding remarks}
\setcounter{equation}{0}
\label{sec-cr}

In this paper we have revisited the classical Whitham approach to the modulation of gravity waves coupled to mean flow and recast these equation in the form of shallow water waves to aid in their interpretation and
analysis. The abstraction of the form of the wave action in this setting is of benefit, which will likely allow for the treatment of Stokes waves when capillarity is considered or flexural terms are considered.  Furthermore the Benjamin-Feir threshold is formulated using these shallow water principles, allowing for more accurate theories than that presented here (such as with further Stokes expansion terms) to be addressed and studied.

At coalescing characteristics the Whitham equations are no longer valid, but
a remodulation with a slower time scale leads to a new modulation equation.
It is shown in \cite{br20}, that conservation of wave action, on the 
slower time scale, and after projection, leads to a modulation equation of
the form
\begin{equation}\label{two-way-bouss}
\mu U_{TT} + \kappa UU_X + \mathscr{K}
U_{XXXX} = 0 \,,
\end{equation}
where $X=\eps x$ and $T=\eps^2t$ are slow space and time scales, $U$ is
 a projection on wavenumber space, and the parameters $\mu$, $\kappa$, and
$\mathscr{K}$ are determined from the averaged Lagrangian. This 
modulation equation is a form of the two-way Boussinesq equation which
has both nonlinearity and dispersion and has a multitude of interesting
solutions.  The story turns out to be far richer than this previous stipulation 
owing to a loss of genuine nonlinearity at the Benjamin-Feir transition for Stokes waves, 
invoking further nonlinear terms into the two-way Boussinesq:
\[
\mu U_{TT}+\alpha_1(U^3)_{XX}+\alpha_2(2UU_T+U_X\partial_X^{-1}U_T)+\mathscr{K}U_{XXXX} = 0
\]
where $\partial_X^{-1}$ denotes the antiderivative with respect to $X$. 
This modulation equation was first derived in \textsc{Ratliff}~\cite{ratliff-thesis}.  Analysis of this equation is more involved, and more interesting,
than the two-way Boussinesq equation, and it will be discussed elsewhere  \cite{rb21}.

One outcome of the study of the water wave problem from this 
Whitham perspective is that it highlights how the approach is unable to capture
the high frequency instabilities reported in other works~\cite{do11}. These non-modulational stabilities
are omitted from these analyses since there are finite-wavelength instabilities as opposed to the long-wavelength
nature of modulation (and related) approaches.

\vspace{0.5cm}

\begin{center}
\hrule height.15 cm
\vspace{.2cm}
--- {\Large\bf Appendix} ---
\vspace{.2cm}
\hrule height.15cm
\end{center}
\vspace{.25cm}

\begin{appendix}

\renewcommand{\theequation}{A-\arabic{equation}}
\section{WMEs for two-phase wavetrains of CNLS}
\label{app-a}
\setcounter{equation}{0}

The CNLS equations (\ref{cnls}) 
are generated by the Lagrangian variational principle
\[
\delta\int_{t_1}^{t_2}\int_{x_1}^{x_2} L(\Psi_1,\Psi_2)\,\rd x\rd t =0\,,
\]
with Lagrangian density
\[
\begin{array}{rcl}
L &=&\displaystyle \ri \left(\overline\Psi_1\frac{\partial\Psi_1}{\partial t}-
\Psi_1\overline{\frac{\partial\Psi_1}{\partial t}}\right) -\alpha_1 \left|\frac{\partial\Psi_1}{\partial x}\right|^2 +\fr \beta_{11}|\Psi_1|^4
+ \beta_{12} |\Psi_1|^2|\Psi_2|^2\\[4mm]
&&\hspace{0.5cm} +\displaystyle \ri \left(\overline\Psi_2\frac{\partial\Psi_2}{\partial t}-
\Psi_2\overline{\frac{\partial\Psi_2}{\partial t}}\right)
-\alpha_2 \left|\frac{\partial\Psi_2}{\partial x}\right|^2 +\fr \beta_{22}|\Psi_2|^4 \,.
\end{array}
\]
Here, a derivation of the WMEs in (\ref{wmes-summary-cnls})
 is sketched following the theory in \textsc{Ratliff}~\cite{r17}.

Substitute the two-phase wavetrain (\ref{2-phase}) into the Lagrangian density
and average over the phase,
\[
\begin{array}{rcl}
\mathscr{L}(\omega_1,k_1,|A_1|^2,\omega_2,k_2,|A_2|^2) &=&\displaystyle
\big(2\omega_1-\alpha_1 k_1^2\big) |A_1|^2  
+\big(2\omega_2-\alpha_2 k_2^2\big) |A_2|^2  \\[4mm]
&&\hspace{.5cm}
+ \fr \beta_{11} |A_1|^4 + \fr \beta_{22} |A_2|^4 +\beta_{12}|A_1|^2|A_2|^2\,.
\end{array}
\]
The derivatives are
\[
\begin{array}{rcl}
\mathscr{L}_{\overline{A_1}} &=& (2\omega_1-\alpha_1 k_1^2)A_1 + \beta_{11}|A_1|^2A_1
+\beta_{12}|A_2|^2A_1\\[2mm]
\mathscr{L}_\omega &=& 2|A_1|^2 \\[2mm]
\mathscr{L}_k &=& -2\alpha_1 k_1 |A_1|^2\,,
\end{array}
\]
and
\[
\begin{array}{rcl}
\mathscr{L}_{\overline{A_2}} &=& (2\omega_2-\alpha_2 k_2^2)A_2 + \beta_{12}|A_1|^2A_2
+\beta_{22}|A_2|^2A_2\\[2mm]
\mathscr{L}_\omega &=& 2|A_2|^2 \\[2mm]
\mathscr{L}_k &=& -2\alpha_2 k_2 |A_2|^2\,.
\end{array}
\]
Solving the first equation of each set gives the nonlinear
dispersion relations (\ref{nonl-disp-cnls}).
Conservation of wave action for each phase is
\[
0 = (\mathscr{L}_{\omega_j})_t-(\mathscr{L}_{k_j})_x =
\big(2|A_j|^2\big)_t + \big(2\alpha_j k_j |A_j|^2\big)_x \,,
\]
or
\[
\big(|A_j|^2\big)_t + \big(\alpha_j k_j |A_j|^2\big)_x = 0 \,.
\]
Noting that $c_g^{(j)} = \alpha_jk_j$, define
\[
h_j = \fr |A_j|^2 \qand u_j = c_g^{(j)} = \alpha_jk_j\,.
\]
Then conservation of wave action becomes
\[
\frac{\partial\ }{\partial t}\left( h_j\right) + 
\frac{\partial\ }{\partial x}\left( u_jh_j\right) = 0 \,,\quad j=1,2\,.
\]
This result confirms the first and third equations in (\ref{wmes-summary-cnls}).

Now look at conservation of waves.  For the $j=1$ component,
\[
\frac{\partial\ }{\partial t}\big( k_1\big) + 
\frac{\partial\ }{\partial x}\big( \omega_1\big) =0\,.
\]
Multiply by $\alpha_1$ and substitute for $\omega_1$ from the nonlinear dispersion relation (\ref{nonl-disp-cnls}),
\[
\frac{\partial\ }{\partial t}\big( \alpha_1 k_1\big) + 
\alpha_1\frac{\partial\ }{\partial x}\left( 
\fr \alpha_1 k_1^2 - \fr\beta_{11}|A_1|^2 - \fr\beta_{12}|A_2|^2 \right) =0\,.
\]
Now substitute for $u_1=\alpha_1k_1$ and $2h_j = |A_j|^2$,
\[
\frac{\partial u_1}{\partial t} + 
u_1\frac{\partial u_1}{\partial x} + g_1'\frac{\partial h_1}{\partial x}
= \alpha_1\beta_{12}\frac{\partial h_2}{\partial x}\,,
\]
with
\[
g_1' = - \alpha_1\beta_{11}\,.
\]
This result confirms the second equation in (\ref{wmes-summary-cnls}).
A similar argument reduces the second component of the conservation of waves,
$(k_2)_t+(\omega_2)_x=0$, to the fourth equation in  (\ref{wmes-summary-cnls}).
This completes the derivation of (\ref{wmes-summary-cnls}).

\end{appendix}

\bibliographystyle{amsplain}

\end{document}